\documentclass[12pt]{iopart}
\usepackage{cite}
\usepackage[usenames,dvipsnames]{xcolor}
\definecolor{ao(english)}{rgb}{0.0, 0.5, 0.0}
\usepackage[breaklinks,colorlinks=true,linkcolor=blue,citecolor=ao(english)]{hyperref}
\usepackage[title,titletoc]{appendix}
\usepackage[english]{babel}
\usepackage[utf8]{inputenc}
\usepackage[OT2,T1]{fontenc}
\usepackage{graphicx}
\usepackage{amssymb,amsfonts,amsthm}%amsmath,
\usepackage{iopams}
\usepackage{bm,bbm}
\usepackage{tikz,stackrel}
\usepackage{pstricks}
\usepackage{pifont} %for ding symbols
\usepackage{changepage}
\usepackage{silence} % silences warnings
\newcommand{\bc}{\begin{center}}
\newcommand{\ec}{\end{center}}
\newcommand{\be}{\begin{equation}}
\newcommand{\ee}{\end{equation}}
\newcommand{\ba}{\begin{eqnarray}}
\newcommand{\ea}{\end{eqnarray}}
\def\bs{\begin{subequations}}
\def\es{\end{subequations}}
\renewcommand{\leq}{\leqslant}
\renewcommand{\geq}{\geqslant}
\def\a{\alpha}
\def\b{\beta}
\def\de{\delta}
\def\g{\gamma}

\def\k{\kappa}

\def\ve{\varepsilon}

\def\t{\tau}  
\def\s{\sigma}

\def\vp{\varphi}
\def\N{\nabla}
\def\cA{\mathcal{A}}

\def\cD{\mathcal{D}}

\def\cF{\mathcal{F}}

\def\cK{\mathcal{K}}
\def\cL{\mathcal{L}}

\def\cO{\mathcal{O}}

\def\cR{\mathcal{R}}

\def\H{{\rm H}}
\def\ds{d_\textsc{s}}
\def\dh{d_\textsc{h}}

\def\p{\partial}
\def\bp{\bar{\partial}}
\def\B{\Box}
\newcommand{\Eq}[1]{(\ref{#1})}

\def\cob{\color{blue}}
\newcommand{\au}[2]{#2 #1}
\newcommand{\ua}[2]{#1 #2}
\newcommand{\book}[5]{\emph{#1} (#3: #2)}
\newcommand{\books}[4]{\emph{#1} (#3: #2)}
\newcommand{\oarX}[1]{\href{http://arxiv.org/abs/#1}{{\ttfamily\cob arXiv:#1}}}
\newcommand{\arX}[1]{\href{http://arxiv.org/abs/#1}{{\ttfamily\cob arXiv:#1}}}
\newcommand{\doin}[6]{\href{http://dx.doi.org/#1}{{\cob {\it #2} #3 {\bf #4} #5}}}
\newcommand{\doinn}[5]{\href{http://dx.doi.org/#1}{{\cob {\it #2} {\bf #3} #4}}}
\newcommand{\doij}[5]{\href{http://dx.doi.org/#1}{{\cob {\it #2} #3(#5)#4}}}

\newcommand{\ndoinn}[5]{\href{#1}{{\cob {\it #2} {\bf #3} #4}}}
\newcommand{\procsinm}[5]{\emph{#1} ed #2 (#4: #3)}
\newcommand{\tia}[1]{#1}
\newcommand{\boxd}[1]{\fbox{$\displaystyle\phantom{\Biggl(}#1\phantom{\Biggl)}$}}

\def\lp{\ell_{\rm Pl}}

\newcounter{listcounter}

\makeatletter
\long\def\@makefntext#1{\parindent 1em\noindent 
 \makebox[1em][l]{\footnotesize\rm$\m@th{^\arabic{footnote}}$}%
 \footnotesize\rm #1}
\def\@makefnmark{\hbox{$^{\arabic{footnote}}\m@th$}}
\def\@thefnmark{\arabic{footnote}}
\makeatother

\begin{document}

\title{Classical and quantum gravity with\\ fractional operators}

\author{Gianluca Calcagni}
\address{Instituto de Estructura de la Materia, CSIC, Serrano 121, 28006 Madrid, Spain}
\ead{g.calcagni@csic.es}
\vspace{10pt}
\begin{indented}
\item[]Original: February 3, 2021. Revision: May 26, 2021
\end{indented}

%\date{February 3, 2021}

\begin{abstract}
Following the same steps made for a scalar field in a parallel publication, we propose a class of perturbative theories of quantum gravity based on fractional operators, where the kinetic operator of the graviton is made of either fractional derivatives or a covariant fractional d'Alembertian. The classical action for each theory is constructed and the equations of motion are derived. Unitarity and renormalizability of theories with a fractional d'Alembertian are also considered. We argue that unitarity and power-counting renormalizability never coexist, although in some cases one-loop unitary and finiteness are possible. One of the theories is unitary and infrared-finite and can serve as a ghost-free model with large-scale modifications of general relativity.
\end{abstract}

% 04.30.-w Gravitational waves
% 04.50.-h Higher-dimensional gravity and other theories of gravity
% 04.60.Bc Phenomenology of quantum gravity
% 05.45.Df Fractals

%\pacs{04.30.-w, 04.50.-h, 04.60.Bc, 05.45.Df}


\centerline{2021 \doinn{10.1088/1361-6382/ac1081}{Class.\ Quantum Grav.}{38}{165005}{2021}; \doinn{10.1088/1361-6382/ac1bea}{Erratum-ibid.}{38}{}{2021} [\arX{2106.15430}]}

%\maketitle

\tableofcontents

\markboth{}{}

%%%%%%%%%%%%%%%%%%%%%%%%%%%%%%%%%%%%%%%%%%%%%%%%%%%%%%%%%%%%%%%%%%%%%%%%%%%%%
%%%%%%%%%%%%%%%%%%%%%%%%%%%%%%%%%%%%%%%%%%%%%%%%%%%%%%%%%%%%%%%%%%%%%%%%%%%%%

\section{Introduction}\label{intro}

Quantum gravity is a broad research field gathering frameworks that combine quantum mechanics and the gravitational force in a consistent way. On one hand, there are approaches that quantize gravity as a fundamental force or they embed it in a unified theory of elementary interactions. In this case, new physics is expected to emerge at short, ultraviolet (UV) scales of order of the Planck length $\lp\sim 10^{-35}\,{\rm m}$. This does not imply that such theories are unobservable; for example, Planckian scales may show up in the sky if quantum corrections were not negligible when cosmological inflation took place. Several instances of this amplification mechanism in quantum gravity exist \cite{Calcagni:2017sdq}. On the other hand, quantum gravity may also indicate an effective field theory with infrared (IR) corrections manifesting themselves at large scales. In this case, one does not necessarily seek to embed these models in a complete fundamental theory and the focus is directed towards phenomenology and IR deviations from general relativity, which may be observable in near-future cosmological observations \cite{Belgacem:2020pdz}.

Some candidates in both the above groups (theories or models with UV or IR corrections) are non-local, i.e., the classical action and equations of motion are characterized by operators $\cF(\B)$ with infinitely many derivatives, where $\B=\N_\mu\N^\mu$ is the covariant Laplace--Beltrami operator or d'Alembertian. The name non-local stems from the fact that such operators can be represented as the convolution of a field $\vp(x)$ with a given integral kernel $F(y-x)$ over all points of spacetime, $\cF(\B)\,\vp(x) = \int\rmd^D y\,F(y-x)\,\vp(y)$ \cite{Pais:1950za}. As is well known, quantum field theory (QFT) is non-local at the quantum level because loop corrections to the bare propagator typically contain non-polynomial functions of momentum. In general, the quantum effective action contains non-local modifications that, in the case of gravity, may leave an imprint in the IR. This IR non-locality is an effective description of a certain regime of a fundamentally local theory. In contrast, UV non-locality refers to theories that are fundamentally non-local already at the classical level. Historically, UV non-locality is as old as IR non-locality. QFT pioneers such as Wataghin \cite{Wataghin:1934ann} and Yukawa \cite{Yuk49,Yukawa:1950eq} considered non-locality as a means to give particles a finite radius and, thus, to cure the infinities of their self-energy \cite{Pau33}. For this reason, non-local operators are also called form factors. Nowadays, fundamental non-locality is invoked for about the same reason (to remove the classical singularities of general relativity and to improve its renormalizability at the quantum level) as well as to preserve unitarity (absence of ghosts and conservation of probability). 

Examples of UV non-local theories are string theory \cite{Zwi09}, non-local quantum gravity with exponential or asymptotically polynomial form factors \cite{Modesto:2011kw,BGKM,Modesto:2017sdr} and multi-fractional QFTs \cite{revmu,Calcagni:2021ipd} (collectively denoted with the label $T_\g$) with fractional operators \cite{frc1,frc2,fra6,frc4,mf0,mf1}, while models with inverse powers of the d'Alembertian realize IR non-locality \cite{Belgacem:2020pdz,Barvinsky:2003kg}. Among these proposals, multi-fractional QFTs with fractional operators are the youngest and the least studied. The purpose of this paper is to initiate a systematic study of the classical and quantum gravitational interaction within this paradigm.

\begin{figure}
\bc
\includegraphics[width=13cm,bb=140 230 595 715,clip=true]{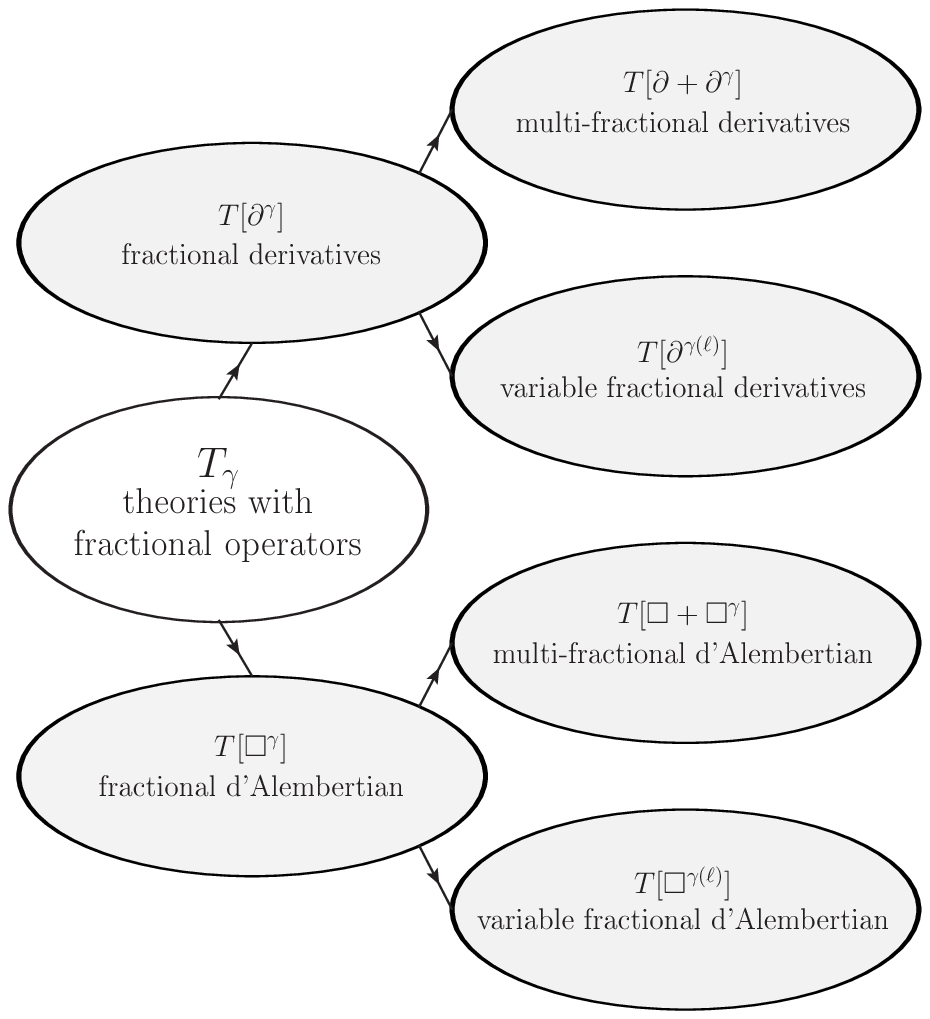}% l b r t
\ec
\caption{\label{fig1} The theories with fractional operators studied in this paper.}
\end{figure}

A first step was made in a companion paper \cite{mf1} that studied the differential structure of the theories $T_\g$ and the classical and quantum properties of a real scalar field living in multi-fractional Minkowski spacetime. There, we classified theories with fractional operators into six types (figure \ref{fig1}). When the non-local form factors are made of derivatives of non-integer order $\g$, ordinary Lorentz symmetry is deformed. One can consider a kinematics with just one type of fractional derivative (theory $T[\p^\g]$) or with fractional derivatives mixed with integer-order derivatives (theory $T[\p+\p^\g]$) or, again, with fractional and integer derivatives distributed on a continuum parametrized by a length scale $\ell$ (theory $T[\p^{\g(\ell)}]$). Among these three possibilities, the first corresponds to a spacetime with fixed non-integer dimension and requires $\g\approx 1$ to recover all the known constraints of the Standard Model of particle physics and of general relativity, while the other two cases correspond to a spacetime with a varying dimension where the Standard Model and general relativity are naturally recovered at certain scales. The same subdivision holds for theories (labeled $T[\B^\g]$, $T[\B+\B^\g]$ and $T[\B^{\g(\ell)}]$) which preserve covariance and where the form factors are non-integer powers of the d'Alembertian.

In \cite{mf1}, we studied the unitarity and renormalizability of scalar QFTs with these operators, developing more in detail the cases with fractional d'Alembertian. Here we will translate those findings to gravity after matching the kinetic term of the graviton to each one of the scalar cases. All the considerations made in \cite{mf1} apply, \emph{mutatis mutandis}, to gravity when treated as a perturbative QFT. We will write down the non-linear action and full equations of motion in all six fractional cases, as well the linearized equations for the graviton. The action of the theory $T[\B+\B^\g]$ is similar to the one for non-local quantum gravity with asymptotically polynomial operators. Both theories are non-local but $T[\B+\B^\g]$ (like all the theories $T_\g$) is characterized by branch cuts, while the other only has simple poles. To summarize the main results at the quantum level, the theory $T[\B^\g]$ with fractional d'Alembertian with $\g>1$ is problematic because its derivatives of non-integer order have the same advantage (suppression of the propagator in the UV) and disadvantage (loss of unitarity \cite{Ste77}) of higher-order derivatives. When $\g<1$, the theory is no longer power-counting renormalizable but it is at least one-loop finite. The theories $T[\B^{\g\approx 1}]$ and $T[\B^{\g(\ell)}]$, where the kinetic term consists of only one fractional operator with, respectively, exponent very close to 1 or with scale-dependent exponent, can be regarded as fundamental quantum theories. Otherwise, if the kinetic term is composed by a d'Alembertian plus a fractional d'Alembertian with $\g<1$ (theory $T[\B+\B^\g]$), then its properties in the UV are those of a standard QFT but, as a payback, it displays modifications dominant in the infrared (IR) without problems of stability, with possible consequences for cosmology.

The plan of the paper is as follows. Classical and quantum gravity with fractional derivatives is discussed in section \ref{tiga}, while the theories with fractional d'Alembertian are studied in section \ref{fradag}. A comparison with other results in the literature of gravity with fractional operators is made in section \ref{compa}. Conclusions and an outlook are in section \ref{conc}. The presentation of the results will be self-contained within certain limits. For all the details concerning multi-fractional spacetimes and the unitarity and renormalizability of the scalar QFTs, we refer the reader to \cite{mf1}.

%%%%%%%%%%%%%%%%%%%%%%%%%%%%%%%%%%%%%%%%%%%%%%%%%%%%%%%%%%%%%%%%%%%%%%%%%%%%%
%%%%%%%%%%%%%%%%%%%%%%%%%%%%%%%%%%%%%%%%%%%%%%%%%%%%%%%%%%%%%%%%%%%%%%%%%%%%%

\section{Gravity with fractional derivatives}\label{tiga}

Having studied in depth the scalar QFT with fractional derivative operators in \cite{mf1}, we turn to the gravitational sector, limiting the discussion to the definition of the action, the classical equations of motion and some of the quantum properties.

The calculation of the equations of motion will be as in standard general relativity except for a non-trivial tensor term $\cO_{\mu\nu}$. Before considering each specific theory $T[\p^\g]$, $T[\p+\p^\g]$ and $T[\p^{\g(\ell)}]$, we write down some variation formul\ae\ for a generic derivative operator $\mathbf{D}$, a generic Levi-Civita connection
\be\label{lecigen}
\mathbf{\Gamma}^\s_{\mu\nu} := \frac12 g^{\rho\s}\left(\mathbf{D}_{\mu} g_{\nu\s}+\mathbf{D}_{\nu} g_{\mu\s}-\mathbf{D}_\s g_{\mu\nu}\right),
\ee
and a generic Ricci tensor and Ricci scalar
\ba
\cR_{\mu\nu}[\mathbf{D}] &:=& {\mathbf{D}}_\s \mathbf{\Gamma}^\s_{\mu\nu}-{\mathbf{D}}_\nu \mathbf{\Gamma}^\s_{\mu\s}+\mathbf{\Gamma}^\tau_{\mu\nu}\mathbf{\Gamma}^\s_{\s\tau}-\mathbf{\Gamma}^\tau_{\mu\s}\mathbf{\Gamma}^\s_{\nu\tau}\,,\label{ricgen}\\
\cR[\mathbf{D}] &:=& g^{\mu\nu}\cR_{\mu\nu}[\mathbf{D}]\,.
\ea
We also define the fractional generalization of a covariant rank-2 tensor,
\be\label{fracov}
\bm{\N}_\s A_{\mu\nu}:=\mathbf{D}_\s A_{\mu\nu}-\mathbf{\Gamma}_{\s\mu}^\rho A_{\rho\nu}-\mathbf{\Gamma}_{\s\nu}^\rho A_{\mu\rho}\,.
\ee
In particular,
\be\label{mecom}
\bm{\N}_\s g_{\mu\nu}=0\,,
\ee
Similarly, for a contravariant vector
\be\label{NAs}
\bm{\N}_\s A^{\mu}=\mathbf{D}_\s A^\mu+\mathbf{\Gamma}_{\s\nu}^\mu A^\nu\,.
\ee

We work in signature $(-,+,\cdots,+)$. Recall that
%\bs
\ba
\de g_{\s\t}&=&-g_{\s\mu}g_{\t\nu}\de g^{\mu\nu}\,,\label{altobasso}\\
\de (g_{\a\mu}g_{\b\nu})A^{\a\b}B^{\mu\nu}&=&-\de (g^{\a\mu}g^{\b\nu})A_{\a\b}B_{\mu\nu}\,,\label{altobasso2}\\
\de\sqrt{|g|} &=& -\frac12\,g_{\mu\nu}\,\sqrt{|g|}\,\delta g^{\mu\nu}\,,\label{desg}
\ea%\es
where $g$ is the determinant of the metric. Then, one has
\ba
\fl \de\mathbf{\Gamma}^\rho_{\mu\nu} &=& \frac12 \de g^{\rho\s}\left(\mathbf{D}_{\mu} g_{\nu\s}+\mathbf{D}_{\nu} g_{\mu\s}-\mathbf{D}_\s g_{\mu\nu}\right)
+ \frac12 g^{\rho\s}\left(\mathbf{D}_{\mu} \de g_{\nu\s}+\mathbf{D}_{\nu} \de g_{\mu\s}-\mathbf{D}_\s \de g_{\mu\nu}\right)\nonumber\\
\fl &\stackrel{\textrm{\tiny \Eq{altobasso}}}{=}&-\frac12 g^{\rho\a}g^{\s\b}\de g_{\a\b}\left(\mathbf{D}_{\mu} g_{\nu\s}+\mathbf{D}_{\nu} g_{\mu\s}-\mathbf{D}_\s g_{\mu\nu}\right)+ \frac12 g^{\rho\s}\left(\mathbf{D}_{\mu} \de g_{\nu\s}+\mathbf{D}_{\nu} \de g_{\mu\s}-\mathbf{D}_\s \de g_{\mu\nu}\right)\nonumber\\
\fl &=&- g^{\rho\a}\de g_{\a\b}\mathbf{\Gamma}^\b_{\mu\nu}+\frac12 g^{\rho\s}\left(\mathbf{D}_{\mu} \de g_{\nu\s}+\mathbf{D}_{\nu} \de g_{\mu\s}-\mathbf{D}_\s \de g_{\mu\nu}\right)\nonumber\\
\fl &=&\frac12 g^{\rho\s}\left(\mathbf{D}_{\mu} \de g_{\nu\s}+\mathbf{D}_{\nu} \de g_{\mu\s}-\mathbf{D}_\s \de g_{\mu\nu}-2\mathbf{\Gamma}^\b_{\mu\nu}\de g_{\s\b}\right)\nonumber\\
\fl &=&\frac12 g^{\rho\s}\left(\mathbf{D}_{\mu} \de g_{\nu\s}-\mathbf{\Gamma}_{\mu\s}^\rho \de g_{\nu\rho}+\mathbf{D}_{\nu} \de g_{\mu\s}-\mathbf{\Gamma}_{\nu\s}^\rho \de g_{\mu\rho}\right.\nonumber\\
\fl &&\qquad\quad\left.-\mathbf{D}_\s \de g_{\mu\nu} + \mathbf{\Gamma}_{\s\mu}^\rho \de g_{\rho\nu}+\mathbf{\Gamma}_{\s\nu}^\rho \de g_{\mu\rho}-2\mathbf{\Gamma}^\b_{\mu\nu}\de g_{\s\b}\right)\nonumber\\
\fl &\stackrel{\textrm{\tiny \Eq{fracov}}}{=}&\frac12 g^{\rho\s}\left(\bm{\N}_\mu \de g_{\nu\s}+\bm{\N}_\nu \de g_{\mu\s}-\bm{\N}_\s \de g_{\mu\nu}\right).\label{lecide}
\ea
The same expression can be inferred by noting that $\de\mathbf{\Gamma}^\rho_{\mu\nu}$ is a tensor, so that its form in a local inertial frame, $\de\mathbf{\Gamma}^\rho_{\mu\nu} = \eta^{\rho\s}(\mathbf{D}_{\mu} \de g_{\nu\s}+\mathbf{D}_{\nu} \de g_{\mu\s}-\mathbf{D}_\s \de g_{\mu\nu})/2$, can be immediately promoted to an arbitrary frame where the Minkowski metric $\eta^{\rho\s}$ is replaced by the generic metric $g^{\rho\s}$ and derivatives are replaced by covariant derivatives.

From definition \Eq{ricgen}, one can check that the variation of the fractional Ricci tensor is
\be\label{decR}
\de \cR_{\mu\nu}[\mathbf{D}]=\bm{\N}_\s\de\mathbf{\Gamma}^\s_{\mu\nu}-\bm{\N}_\nu\de\mathbf{\Gamma}^\rho_{\mu\rho}\,,
\ee
so that
\ba
\de g^{\mu\nu}\cO_{\mu\nu} &:=& g^{\mu\nu}\de\cR_{\mu\nu}[\mathbf{D}]\nonumber\\
&\stackrel{\textrm{\tiny \Eq{mecom}}}{=}&\bm{\N}_\s\left(g^{\mu\nu}\de\mathbf{\Gamma}^\s_{\mu\nu}-g^{\mu\s}\de\mathbf{\Gamma}^\rho_{\mu\rho}\right)\nonumber\\
&=:& \bm{\N}_\s \cA^\s\label{Omn1}\\
&\stackrel{\textrm{\tiny \Eq{NAs}}}{=}&  \mathbf{D}_\s\cA^\s+\mathbf{\Gamma}_{\s\nu}^\s\cA^\nu\nonumber\\
&\stackrel{\textrm{\tiny \Eq{lecigen}}}{=}&  \mathbf{D}_\s\cA^\s+\frac12 g^{\rho\s}\mathbf{D}_\nu g_{\rho\s}\cA^\nu\,,\label{Omn2}
\ea
where in the last line we used $\mathbf{\Gamma}^\s_{\s\nu}= g^{\rho\s}\mathbf{D}_\nu g_{\rho\s}/2$.

In ordinary calculus, the tensor $\cO_{\mu\nu}$ would be identically zero in the absence of boundary or on any boundary where $\de g^{\mu\nu}=0$, or it could be cancelled by the York--Gibbons--Hawking boundary term. The argument would be the following. For any matrix $M$, the logarithm-trace formula $\ln(\det M)={\rm tr}(\ln M)$ holds. Taking the first derivative on both sides, $(\det M)^{-1}\p \det M={\rm tr}(M^{-1}\p M)$. For the metric $M=g_{\mu\nu}$, in the sense of variations this reads $\de g=g\,g^{\mu\nu}\de g_{\mu\nu}=-g\,g_{\mu\nu}\de g^{\mu\nu}$, which yields \Eq{desg}, while in the sense of spacetime derivatives it implies
\be\label{sqrt0}
\frac{1}{\sqrt{|g|}}\p_\s \sqrt{|g|}=\frac12\p_\s\ln g=\frac12 g^{\mu\nu}\p_\s g_{\mu\nu}\,.
\ee
Therefore, in the ordinary case the integrand in \Eq{Omn2} would be a total derivative,
\be\label{sqrt}
\N_\s \cA^\s = \p_\s\cA^\s+\frac{1}{\sqrt{|g|}}\p_\s \sqrt{|g|}\cA^\nu=\frac{1}{\sqrt{|g|}}\p_\s \left(\sqrt{|g|}\cA^\nu\right)\,,
\ee
which vanishes at infinity:
\be
\fl \int\rmd^Dx\,\sqrt{|g|}\,\de g^{\mu\nu}\cO_{\mu\nu}^{\mathbf{D}=\p} = \int\rmd^Dx\,\sqrt{|g|}\,\N_\s \cA^\s=\int\rmd^Dx\,\p_\s \left(\sqrt{|g|}\cA^\nu\right)=0\,,
\ee
so that $\cO_{\mu\nu}^{\mathbf{D}=\p}=0$ in the Einstein equations. However, there is no simple fractional analogue of \Eq{sqrt0} and \Eq{sqrt} because the fractional derivative of a composite function $f[g(x)]$ stems from a highly non-trivial Leibniz rule \cite{Pod99}. It may still be possible that, once the rule for the mixed multi-fractional derivative of a composite function was found, and taking a non-trivial multi-fractional integration measure $\rmd^Dx\to \rmd^Dx\,v(x)$ with scaling matching the one of the fractional derivatives \cite{revmu,Calcagni:2021ipd}, one could obtain a simple or even vanishing expression for the operator $\cO_{\mu\nu}$. We will not solve this mathematical problem here.

%%%%%%%%%%%%%%%%%%%%%%%%%%%%%%%%%%%%%%%%%%%%%%%%%%%%%%%%%%%%%%%%%%%%%%%%%%%%%

\subsection{Theory \texorpdfstring{$T[\p^\g]$}{Tpg}}

The gravitational theory with multi-fractional derivatives can be written down in a straightforward manner noting that, according to the paradigm \cite{revmu,Calcagni:2021ipd}, the whole integro-differential structure is supplanted by a multi-fractional structure, i.e., spacetime is endowed with one or more fundamental length scales. This means, in particular, that the candidate action should resemble the Einstein--Hilbert action where the integration measure $\rmd^Dx$ is the same, where $D$ is the topological dimension of spacetime ($D=4$ in the physical case), and all ordinary derivatives are replaced by fractional or multi-fractional derivatives. In \cite{mf0,mf1}, we defined fractional derivative operators $\cD_{\pm\mu}^\g$ mixing Liouville and Weyl fractional derivatives in such a way that one can consider spacetime coordinates of any sign:
\be\label{tD+-}
\cD^\g_{\pm\mu}:=\frac12\left({}_\infty\p^\g_\mu\pm{}_\infty\bp^\g_\mu\right),
\ee
where, omitting the coordinate index $\mu$,
\ba
\fl {}_\infty\p^\g f(x) &:=& \frac{1}{\Gamma(m-\g)}\int_{-\infty}^{x}\, \frac{\rmd x'}{(x-x')^{\g+1-m}}\p_{x'}^m f(x')\,,\qquad m-1\leq \g<m\,,\\
\fl {}_\infty\bar\p^\g f(x) &:=& \frac{1}{\Gamma(m-\g)}\int_{x}^{+\infty}\, \frac{\rmd x'}{(x'-x)^{\g+1-m}}\p_{x'}^m f(x')\,,\qquad m-1\leq \g<m\,,
\ea
are, respectively, the Liouville and the Weyl derivative \cite{MR,Pod99,SKM,KST} for each spacetime direction and $m=1,2,\dots$. Their properties can be found in \cite{mf1}. Here we only recall the composition rule
\be\label{comD}
\fl \cD^\g_\pm\cD^\b_\pm = \frac{1\pm 1}{2}\cD_\pm^{\g+\b}\pm\frac14({}_{\infty}\p^\g\,{}_{\infty}\bp^\b+{}_{\infty}\bp^\g\,{}_{\infty}\p^\b-{}_{\infty}\p^{\g+\b}-{}_{\infty}\bp^{\g+\b})\,,%2c\bar c\cD^{\g+\b} +c\bar c({}_{\infty}\p^\g\,{}_{\infty}\bp^\b+{}_{\infty}\bp^\g\,{}_{\infty}\p^\b)
\ee
and the Leibniz rule
\be\label{leruD}
\cD^\g_\pm(fg)=\sum_{j=0}^{+\infty}\frac{\Gamma(\g+1)}{\Gamma(\g-j+1)\Gamma(j+1)} (\p^j f)(\cD^{\g-j}_\pm g)\,.%,\qquad \binom{\g}{j}=\frac{\Gamma(1+\g)}{\Gamma(\g-j+1)\Gamma(j+1)}\,.
\ee

The operators $\cD^\g_\pm$ generalize derivatives of, respectively, even order $\g=2n$ and odd order $\g=2n+1$. In the scalar-field case, we are at liberty of choosing between the kinetic terms $\phi\cD^{2\g}_+\phi$ and $\cD^\g_-\phi\cD^\g_-\phi$ to define the action. The two choices are inequivalent because mixed fractional derivatives do not compose trivially, $\cD_\pm^\g\cD_\pm^\g\neq \cD_\pm^{2\g}$. In the gravitational case, we do not have such an arbitrariness because the equivalent of first-order derivatives appear isolated from one another. Therefore, we must select $\cD_-^\g$ or its multi-fractional generalization 
\be\label{multider}
\cD_{-\mu}:=\sum_\g u_\g\cD^\g_{-\mu}\,,
\ee
where $u_\g=u_\g(\ell_i)$ are dimensionful coefficients depending on one or more fundamental length scales $\ell_1$, $\ell_2$, \dots.

From now on, we omit the subscript $-$. The fractional Levi-Civita connection, Riemann tensor, Ricci tensor and Ricci scalar feature the fractional derivatives $\cD_\mu^\g$ given by \Eq{tD+-} with the $-$ sign choice:
\ba
\tilde\Gamma^\rho_{\mu\nu} &:=& \frac12 g^{\rho\s}\left(\cD_{\mu}^\g g_{\nu\s}+\cD_{\nu}^\g g_{\mu\s}-\cD_\s^\g g_{\mu\nu}\right),\label{leci}\\
R^{(\g)\rho}_{~~~~\mu\sigma\nu}&:=& \cD^\g_\sigma \tilde\Gamma^\rho_{\mu\nu}-\cD^\g_\nu \tilde\Gamma^\rho_{\mu\sigma}+\tilde\Gamma^\tau_{\mu\nu}\tilde\Gamma^\rho_{\sigma\tau}-\tilde\Gamma^\tau_{\mu\sigma}\tilde\Gamma^\rho_{\nu\tau}\,,\\
R_{\mu\nu}^{(\g)}&:=& R^{(\g)\s}_{~~~~\mu\s\nu}\,,\qquad R^{(\g)} := g^{\mu\nu}R_{\mu\nu}^{(\g)}\,.\label{ricg}
\ea
Note that this and the following are metric theories with an added non-dynamical structure that, on one hand, alters the differential structure of the manifold \cite{frc1,frc2} and, on the other hand, implies that covariance and diffeomorphism invariance are no longer equivalent \cite{frc11}. As far as the differential structure is concerned, it is well-known that the commutator of two covariant derivatives acting on a vector measures how the parallel transport of the vector differs if we first transport it one way and then the other, or vice versa. In a flat spacetime, the commutator is zero and the ordering of the directions does not matter. In ordinary curved spacetime, the difference is proportional to the Riemann tensor, $[\N_\mu,\N_\nu]A^\rho = R^{\rho}_{\ \lambda \mu\nu} A^\lambda$ (see section \ref{tppgs} for definitions); the right-hand side expresses the fact that these manifolds are completely characterized by a metric structure. In curved spacetime for the theory $T[\p^\g]$, the fractional covariant derivatives
\be\label{NAst}
\tilde\N_\s A^{\mu}:=\cD_\s^\g A^\mu+\tilde\Gamma_{\s\nu}^\mu A^\nu\,,\qquad \tilde\N_\s A_{\mu}:=\cD_\s^\g A_\mu-\tilde\Gamma_{\s\mu}^\nu A_\nu\,,
\ee
still do not commute but the difference is no longer proportional to the fractional Riemann tensor:
\ba
\fl [\tilde\N_\mu,\tilde\N_\nu]A^\rho &=& R^{(\g)\rho}_{~~~~\lambda \mu\nu} A^\lambda+\sum_{j=1}^{+\infty}\frac{\Gamma(\g+1)}{\Gamma(\g-j+1)\Gamma(j+1)} \p^j A^\lambda (\cD^{\g-j}_\mu\tilde\Gamma_{\nu\lambda}^\rho - \cD^{\g-j}_\nu\tilde\Gamma_{\mu\lambda}^\rho)\nonumber\\
\fl &&-(\tilde\Gamma_{\nu\lambda}^\rho\cD_\mu^\g-\tilde\Gamma_{\mu\lambda}^\rho\cD_\nu^\g) A^\lambda\,,
\ea
where we used \Eq{leruD}. The extra terms show explicitly that the manifold is endowed with a non-metric structure independent of the metric one. The complicated Leibniz rule is, in fact, one of the reasons why we will soon shift our attention from theories with fractional derivatives to theories with fractional d'Alembertian.

Regarding symmetries, ordinary diffeomorphism invariance is deformed and local inertial frames are described by fractional Lorentz transformations \cite{frc2}. It would be interesting to study how diffeomorphism invariance is deformed by computing the hypersurface-deformation algebra of the scalar and vector constraints in Hamiltonian formalism, as done in \cite{CaRo1} for multi-fractional theories with integer-order derivatives.

The gravitational action is
\be\label{actg0}
\boxd{S=\frac{1}{2\k^2}\int\rmd^Dx\,\sqrt{|g|}\,\left[R^{(\g)}-2\Lambda\right],}
\ee
where $\k^2=8\pi G$ is proportional to Newton's constant and $\Lambda$ is a cosmological constant. 

Adding a matter action $S_{\rm m}$ to \Eq{actg0} and applying equations \Eq{lecide} and \Eq{Omn2}, since 
\be
\frac{1}{\sqrt{|g|}}\de(\sqrt{|g|}\cR)\stackrel{\textrm{\tiny \Eq{desg}}}{=} \de g^{\mu\nu}\left(\cR_{\mu\nu}-\frac12\,g_{\mu\nu}\cR\right)+g^{\mu\nu}\de\cR_{\mu\nu}\,,
\ee
one obtains the fractional Einstein equations
\be\label{eomukfix}
R_{\mu\nu}^{(\g)}-\frac12 g_{\mu\nu}R^{(\g)}+\Lambda g_{\mu\nu}+\cO_{\mu\nu}=\k^2T_{\mu\nu}\,,
\ee
where
\be
T_{\mu\nu}:=-\frac{2}{\sqrt{|g|}}\frac{\de S_{\rm m}}{\de g^{\mu\nu}}
\ee
is the matter energy-momentum tensor.

Taken as a physical theory, $T[\p^\g]$ would be such that $\g=1-\ve\lesssim 1$ in order to respect all gravitational phenomenology without violating any experimental constraint. However, we have no theoretical reason justifying such a fine tuning of a $\g$ very close to, but different from, one. Therefore, we regard \Eq{actg0} as a starting point to understand the next theories rather than a physical model by itself.

%%%%%%%%%%%%%%%%%%%%%%%%%%%%%%%%%%%%%%%%%%%%%%%%%%%%%%%%%%%%%%%%%%%%%%%%%%%%%

\subsection{Theory \texorpdfstring{$T[\p+\p^\g]$}{Tppg}}\label{tppgs}

The theory with multi-fractional derivatives can be defined in two ways, which we might call naive and natural. The naive one involves a fractional and an ordinary sector independently added to the action. The fractional sector corresponds to the connection $\tilde\Gamma^\rho_{\mu\nu}$ and Ricci tensor $R_{\mu\nu}^{(\g)}$ of the theory $T[\p^\g]$, while the ordinary sector is described by the usual Levi-Civita connection and curvature tensors:
%\bs\label{tutto}
\ba
\hspace{-1.5cm}&&
\Gamma^\rho_{\mu\nu} := \frac12 g^{\rho\s}\left(\p_\mu g_{\nu\s}+\p_\nu g_{\mu\s}-\p_\s g_{\mu\nu}\right),\label{lecico}\\
\hspace{-1.5cm}&&\textrm{Riemann tensor:}\qquad R^\rho_{~\mu\sigma\nu}:= \p_\sigma \Gamma^\rho_{\mu\nu}-\p_\nu \Gamma^\rho_{\mu\sigma}+\Gamma^\tau_{\mu\nu}\Gamma^\rho_{\sigma\tau}-\Gamma^\tau_{\mu\sigma}\Gamma^\rho_{\nu\tau}\,,\label{rite}\\
\hspace{-1.5cm}&&\textrm{Ricci tensor:}\hspace{1.3cm} R_{\mu\nu}:= R^\rho_{~\mu\rho\nu}\,,\\
\hspace{-1.5cm}&&\textrm{Ricci scalar:}\hspace{1.35cm} R:= R_{\mu\nu}g^{\mu\nu}\,,\\
\hspace{-1.5cm}&&\textrm{Einstein tensor:}\hspace{.8cm} G_{\mu\nu}:= R_{\mu\nu}-\frac12g_{\mu\nu}R\,.\label{Eiten}
\ea%\es
Then, the action is \Eq{actg0} plus the Einstein--Hilbert action:
\be\label{actg2}
\boxd{S=\frac{1}{2\k^2}\int\rmd^Dx\,\sqrt{|g|}\,\left[R+\ell_*^{2(\g-1)}R^{(\g)}-2\Lambda\right].}
\ee
%This formulation has the limited advantage that $R$ and $R^{(\g)}$ are separately invariant under, respectively, ordinary and fractional Lorentz transformations.
The equations of motion are
\be\label{eomuk1}
\left[R_{\mu\nu}+R_{\mu\nu}^{(\g)}\right]-\frac12 g_{\mu\nu}\!\left[R+R^{(\g)}\right]+\Lambda g_{\mu\nu}+\cO_{\mu\nu}=\k^2T_{\mu\nu}\,.
\ee

In contrast, the natural definition of $T[\p+\p^\g]$ realizes multi-scaling within each derivative instead of at the level of action operators. Defining the fractional Levi-Civita connection with the multi-fractional derivatives \Eq{multider} $\cD_-$ (with omitted $-$ subscript),
\be
\bar\Gamma^\rho_{\mu\nu}:= \frac12 g^{\rho\s}\left(\cD_{\mu} g_{\nu\s}+\cD_{\nu} g_{\mu\s}-\cD_\s g_{\mu\nu}\right),\label{leci0}
\ee
we can construct the covariant derivatives \Eq{fracov} and \Eq{NAs}. By reverse engineering, one should be able to determine the transformation laws of vectors and tensors under coordinate transformations in the multi-fractional geometry defined by fractional calculus. Note that angles and lengths are not parallel transported by the Levi-Civita connection and the metric is not covariantly constant, $\N_\s g_{\mu\nu}\neq 0$. Still, there is the fractional notion \Eq{mecom} of metric compatibility,
\be\label{mecombar}
\bar\N_\s g_{\mu\nu}=0\,,
\ee
which, just like the standard one, allows the Minkowski metric in local inertial frames because constants are in the kernel of the fractional derivatives we are using if we further impose \cite{mf1}
\be\label{g01}
0<\g<1\,.
\ee
Then, the Minkowski metric is the only admissible one in local frames, since the kernel of the fractional derivatives becomes trivial. Violation of integer-order metric compatibility is common in multi-fractional spacetimes, in particular, in the theories $T_v$ with weighted derivatives and $T_q$ with $q$-derivatives \cite{frc11}.

The multi-fractional generalization of the Ricci tensor is
\be\label{ricg0}
\cR_{\mu\nu}:= \cD_\s\bar\Gamma^\s_{\mu\nu}-\cD_\nu\bar\Gamma^\s_{\mu\s}+\bar\Gamma^\tau_{\mu\nu}\bar\Gamma^\s_{\s\tau}-\bar\Gamma^\tau_{\mu\s}\bar\Gamma^\s_{\nu\tau}\,,
\ee
while the Ricci scalar is $\cR:=g^{\mu\nu}\cR_{\mu\nu}$. The gravitational action in $D$ topological dimensions reads
\be\label{actg1}
\boxd{S=\frac{1}{2\k^2}\int\rmd^Dx\,\sqrt{|g|}\,\left(\cR-2\Lambda\right).}
\ee
Due to the multi-scaling of the derivatives, at the level of local inertial frames there is no simple analogue of fractional Lorentz transformations \cite{frc2}. Also, contrary to the naive version \Eq{actg2}, the action \Eq{actg1} hides mixed terms of derivative order $1+\g$.

Variation with respect to the metric of the action \Eq{actg1} plus a matter action $S_{\rm m}$ yields the multi-fractional generalization of Einstein's equations
\be\label{eomuk0}
\cR_{\mu\nu}-\frac12 g_{\mu\nu}\cR+\Lambda g_{\mu\nu}+\cO_{\mu\nu}=\k^2T_{\mu\nu}\,.
\ee

%%%%%%%%%%%%%%%%%%%%%%%%%%%%%%%%%%%%%%%%%%%%%%%%%%%%%%%%%%%%%%%%%%%%%%%%%%%%%

\subsection{Theory \texorpdfstring{$T[\p^{\g(\ell)}]$}{Tpgell}}\label{tigall}

This is the analogue of the action \Eq{actg0} with variable-order fractional derivatives $\cD_\mu^{\g(\ell)}$:
\be\label{actg3}
\boxd{S=\frac{1}{2\k^2\ell_*}\int_0^{+\infty}\rmd\ell\,\tau(\ell)\int\rmd^Dx\,\sqrt{|g|}\,\left[R^{(\g(\ell))}-2\Lambda\right].}
\ee
The theory does not have unnecessary mixed-order terms and may allow for a relatively simple variable-order form of fractional Lorentz invariance in local frames \cite[section 2.3]{frc2}. The equations of motion stemming from \Eq{actg3} are \Eq{eomukfix} with the replacement $\g\to\g(\ell)$, without integration on $\ell$. In fact, in a unified theory also the right-hand side is made of matter fields with multi-fractional dynamics, whose action $S_{\rm m}$ also contains fractional exponents $\g(\ell)$ and the $\ell$ integration as in \Eq{actg3}, with same weight. Therefore, the equality in \Eq{eomukfix} must hold for all $\ell$ in the integration range. In a compact notation,
\be\label{eomuk2}
R_{\mu\nu}^{(\g(\ell))}-\frac12 g_{\mu\nu}R^{(\g(\ell))}+\Lambda g_{\mu\nu}+\cO_{\mu\nu}^{(\ell)}=\k^2T_{\mu\nu}^{(\ell)}\,.
\ee

Of the three versions of gravity with multi-fractional derivatives, the easiest but also the least elegant is perhaps \Eq{actg2}, while \Eq{actg3} can keep some simplicity and admit a richer symmetry structure. All three versions share the difficulty of the complicated composition rule \Eq{comD} and Leibniz rule \Eq{leruD} for fractional derivatives, that make the derivation and manipulation of the equations of motion tricky, especially in the calculation of the tensor $\cO_{\mu\nu}$.

%%%%%%%%%%%%%%%%%%%%%%%%%%%%%%%%%%%%%%%%%%%%%%%%%%%%%%%%%%%%%%%%%%%%%%%%%%%%%
%%%%%%%%%%%%%%%%%%%%%%%%%%%%%%%%%%%%%%%%%%%%%%%%%%%%%%%%%%%%%%%%%%%%%%%%%%%%%

\section{Gravity with fractional d'Alembertian}\label{fradag}

In this section, we construct a covariant gravitational theory with fractional Laplace--Beltrami operators with the same quantum properties discovered for the scalar QFT with the same kinetic term \cite{mf1}. %The requirement of diffeomorphism invariance will restrict the possible choices of derivative operators in the action even more than for the scalar field.

Just like in the scalar-field case we proposed the fractional d'Alembertian as a way to preserve ordinary Lorentz invariance, so do we wish to preserve ordinary diffeomorphism invariance and local ordinary Lorentz invariance in the gravity case. The action we look for will then be made of some functions $\cF_i(\B)$ of the covariant d'Alembertian acting on the standard curvature tensors. Spacetime geometry is described by the usual Levi-Civita connection and curvature tensors \Eq{lecico}--\Eq{Eiten}. Angles and lengths are parallel transported and the compatibility equation $\N_\s g_{\mu\nu}=0$ holds as usual.

Our starting point is the generic gravitational action
\be
\fl 
S=%\frac{1}{2\k^2}\int\rmd^Dx\,\sqrt{|g|}\,\cL\nonumber\\
%\hspace{-.9cm}&=&
\frac{1}{2\k^2}\!\int\!\rmd^Dx\,\sqrt{|g|}\,\left[R-2\Lambda+R\cF_1(\B)\,R+G_{\mu\nu}\cF_2(\B)\,R^{\mu\nu}+R_{\mu\nu\rho\s}\cF_3(\B)\,R^{\mu\nu\rho\s}\right],\label{act4}
\ee
plus a matter action $S_{\rm m}$, where the Ricci tensor and Ricci scalar are defined by the ordinary expressions \Eq{lecico}--\Eq{Eiten}. This action is written in the so-called Einstein basis (Riemann tensor, Ricci tensor and Ricci scalar) but one could also adopt the Weyl basis, where the next-to-last term disappears and the Riemann tensor in the last term is replaced by the Weyl tensor.

The Einstein--Hilbert term $R$ is necessary to recover the limit of general relativity, while the second-order curvature invariants are the simplest non-trivial covariant operators where one can insert the functions $\cF_i(\B)$.%, which in our case will be of the form $\cF_i(\B)\propto(-\B)^{\g-2}$, as we will prove shortly. This already gives us valuable information: If we preserve diffeomorphism invariance and work with covariant operators, the multi-fractional kinetic term of the graviton will necessarily be of the type $\cK\sim\B+\B^\g$ (eq.\ \Eq{Kg} with $m=0$), thus excluding the variable-order operator $\sim \B^{\g(\ell)}$ (eq.\ \Eq{Kell} with $m=0$) with extra integration on the scale. In turn, this will severely restrict the quantum properties of the theory because one will no longer be able to fix the ``unitarity vs.\ renormalizability issue'' in the ways proposed in section \ref{rangega}.

The equations of motion from the action \Eq{act4} can be found in \cite{Koshelev:2013lfm,BCKM} for all $\cF_i\neq 0$ and a series representation of the variation of non-local operators, and in \cite{Calcagni:2018lyd} for $\cF_1=0=\cF_3$ and an integral representation of the variation of non-local operators. The integral representation will be very convenient in the case of fractional operators, so that in the following we extend the calculation of \cite{Calcagni:2018lyd} to $\cF_1\neq 0$, although we will check \emph{a posteriori} that this generalization is not necessary. We still keep $\cF_3=0$ for simplicity, since this term does not play any role in the unitarity of the theory.

To derive the equations of motion of the theory \Eq{act4} with $\cF_3=0$, we make use of basic variation formul\ae\ \Eq{altobasso}, \Eq{altobasso2}, \Eq{desg} and
%\bs
\ba
\de R_{\mu\nu}	 &=& \N^\a\N_{(\mu}\de g_{\nu)\a}-\frac12\left[\B\de g_{\mu\nu}+g^{\a\b}\N_{(\mu}\N_{\nu)}\de g_{\a\b}\right]\,,\label{dRmn}\\
\delta R         &=& \de g^{\mu\nu}R_{\mu\nu} +g^{\mu\nu}\de R_{\mu\nu}\stackbin[\textrm{\tiny \Eq{altobasso}}]{\textrm{\tiny \Eq{dRmn}}}{=} (R_{\mu\nu}+g_{\mu\nu}\,\B -\N_\mu\N_\nu)\,\de g^{\mu\nu}\,,\label{dRg}
\ea%\es
where $A_{(\mu}B_{\nu)}:=(A_\mu B_\nu+A_\nu B_\mu)/2$. The variation of the action \Eq{act4} with $\cF_1=0=\cF_3$ can be found step by step in \cite[Appendix E]{Calcagni:2018lyd}:\footnote{We correct a minor typo (a missing $\de g^{\mu\nu}$ in the penultimate term) of (E.4) of the published version of \cite{Calcagni:2018lyd} and also in the last line of (D.13) of the same paper (the $\a\b$ indices in $\Theta_{\mu\nu}$ should be bottom and up instead of top and down). These typos have been fixed in the arXiv version of \cite{Calcagni:2018lyd}.}
\ba
\fl \frac{2\kappa^2}{\sqrt{|g|}}\left.\frac{\de (\sqrt{|g|}\,\cL)}{\de g^{\mu\nu}}\right|_{{\scriptstyle \cF_1=0}\atop {\scriptstyle \cF_3=0}} &=&  G_{\mu\nu}+\Lambda g_{\mu\nu} - \frac{1}{2} g_{\mu\nu} \, G_{\s\t} \cF_2 R^{\s\t}+2G^\s_{\ \mu} \cF_2  G_{\nu\s}\nonumber\\
\fl && +\cF_2\B G_{\mu\nu}+g_{\mu\nu}\N^\s\N^\t\cF_2 G_{\s\t}-2\N^\s\N_{\mu}\cF_2 G_{\nu\s}\nonumber\\
\fl &&+\frac12(G_{\mu\nu} \cF_2 R+R\cF_2 G_{\mu\nu})+\Theta_{\mu\nu}(R_{\a\b},G^{\a\b})+O(\N)\,,\label{c1eq}
\ea
where $O(\N)$ are total derivative terms and we used definition \Eq{thethe}. Here we only need to add to this equation the variation of the $\cF_1$ term:
\ba
\fl \frac{1}{\sqrt{|g|}}\de\left(\sqrt{|g|}R\cF_1R\right) &\stackrel{\textrm{\tiny \Eq{desg}}}{=}& -\frac12\de g^{\mu\nu}g_{\mu\nu}R\cF_1R+2\de R\cF_1 R+R \de\cF_1 R+O(\N)\nonumber\\
\fl &\stackbin[\textrm{\tiny \Eq{dRg}}]{\textrm{\tiny \Eq{thethe}}}{=}& -\frac12\de g^{\mu\nu}g_{\mu\nu}R\cF_1R+2\de g^{\mu\nu}(R_{\mu\nu}+g_{\mu\nu}\,\B -\N_\mu\N_\nu)\cF_1 R\nonumber\\
\fl &&+\de g^{\mu\nu}\vartheta_{\mu\nu}(R,R)+O(\N)\,,\label{c2eq}
\ea
where the variation symbol $\de$ always and only applies to the first operator on its right and
\be\label{thethe}
\vartheta_{\mu\nu}(R,R):=R\frac{\de\cF_1}{\de g^{\mu\nu}}R\,,\qquad \Theta_{\mu\nu}(R_{\a\b},G^{\a\b}):=G_{\rho\s}\frac{\de\cF_2}{\de g^{\mu\nu}}R^{\rho\s}\,.
\ee
Combining \Eq{c1eq} and \Eq{c2eq}, we obtain 
\ba
\fl \k^2 T_{\mu\nu} &=& (1+\cF_2\B) G_{\mu\nu}+\Lambda g_{\mu\nu}\nonumber\\
\fl &&+2[g_{\mu\nu}\B-\N_{(\mu}\N_{\nu)}]\cF_1R+g_{\mu\nu}\N^\s\N^\t\cF_2 G_{\s\t} -2\N^\s\N_{(\mu}\cF_2 G_{\nu)\s}\nonumber\\
\fl &&+(G_{\mu\nu}+R_{\mu\nu})\cF_1 R-\frac{1}{2} g_{\mu\nu}G_{\s\t}\cF_2 R^{\s\t}+2G^\s_{\ (\mu} \cF_2  G_{\nu)\s}+\frac12(G_{\mu\nu} \cF_2  R+R\cF_2 G_{\mu\nu})\nonumber\\
%&&-\frac{1}{2} g_{\mu\nu}\left(R\cF_1 R+G_{\s\t}\cF_2 R^{\s\t}\right)+2\left[R_{\mu\nu}\cF_1 R+G^\s_{\ (\mu} \cF_2  G_{\nu)\s}\right]+\frac12(G_{\mu\nu} \cF_2  R+R\cF_2 G_{\mu\nu})\nonumber\\
\fl &&+\vartheta_{\mu\nu}(R,R)+\Theta_{\mu\nu}(R_{\a\b},G^{\a\b})\,.\label{EinEq1}
\ea

At this point, the choice of form factors $\cF_{1,2}$ determines the theory and, in particular, the terms \Eq{thethe} and the kinetic operator $\cK$ of the graviton. Split the metric into Minkowski background and a perturbation,
\be
g_{\mu\nu}=\eta_{\mu\nu}+h_{\mu\nu}\,.
\ee
Working in the transverse-traceless gauge
\be
\p^\mu h_{\mu\nu}=0=h_{\mu}^{\mu}\,,\label{onshellgrav}
\ee
which can always be selected in any covariant theory on a $D$-dimensional Minkowski background \cite{MTW,Calcagni:2017sdq}, the on-shell graviton is the transverse traceless part $h_{ij}$ of the perturbation. Finding the linearized equation of motion for $h_{\mu\nu}$ in vacuum is easy, since all second-order curvature terms in \Eq{EinEq1} vanish on Minkowski background. The linearization of the Ricci tensor can be read off from \Eq{dRmn} with $\de g_{\mu\nu}=h_{\mu\nu}$, so that by virtue of \Eq{onshellgrav} $\de^{(1)}R_{\mu\nu}= -\B h_{\mu\nu}/2$ and $\de^{(1)} R=0$. Setting $\Lambda=0$ in \Eq{EinEq1}, the modified wave equation for the graviton in vacuum is
\be\label{waveq}
[1+\cF_2(\B)\,\B]\B h_{\mu\nu}=0\,,
\ee
where $\B=\B_\eta$ is the d'Alembertian in Minkowski spacetime. Unitarity can be evinced from the choice of kinetic term $\cK(\B)=\B+\cF_2(\B)\,\B^2$, which does not involve $\cF_1$. Therefore, without loss of generality we can set
\be
\cF_1=0\,.
\ee

%%%%%%%%%%%%%%%%%%%%%%%%%%%%%%%%%%%%%%%%%%%%%%%%%%%%%%%%%%%%%%%%%%%%%%%%%%%%%

\subsection{Theory \texorpdfstring{$T[\B^\g]$}{TBg}}

In this theory, the graviton kinetic term is a single fractional d'Alembertian, which stems from the form factor
\be\label{F22fix}
\cF_2(\B)=\frac{(-\ell_*^2\B)^{\g-1}-1}{\B}\qquad\Longrightarrow\qquad (-\B)^\g\, h_{\mu\nu}=0\,.
\ee
The action and equations of motion for the theory with such form factor are
\be
\boxd{S=\frac{1}{2\k^2}\int\rmd^Dx\,\sqrt{|g|}\,\left[R-2\Lambda+ G_{\mu\nu}\frac{(-\ell_*^2\B)^{\g-1}-1}{\B}\,R^{\mu\nu}\right],\label{act4fin2}}
\ee
and
\ba
\fl \k^2 T_{\mu\nu} &=& (-\ell_*^2\B)^{\g-1} G_{\mu\nu}+\Lambda g_{\mu\nu}\nonumber\\
\fl &&+g_{\mu\nu}\N^\s\N^\t\frac{(-\ell_*^2\B)^{\g-1}-1}{\B} G_{\s\t} -2\N^\s\N_{(\mu}\frac{(-\ell_*^2\B)^{\g-1}-1}{\B} G_{\nu)\s}\nonumber\\
\fl &&-\frac{1}{2} g_{\mu\nu}G_{\s\t}\frac{(-\ell_*^2\B)^{\g-1}-1}{\B} R^{\s\t}+2G^\s_{\ (\mu} \frac{(-\ell_*^2\B)^{\g-1}-1}{\B} G_{\nu)\s}\nonumber\\
\fl &&+\frac12\left[G_{\mu\nu}\frac{(-\ell_*^2\B)^{\g-1}-1}{\B}  R+R\frac{(-\ell_*^2\B)^{\g-1}-1}{\B} G_{\mu\nu}\right]+\Theta_{\mu\nu}(R_{\a\b},G^{\a\b})\,,\label{EinEq1fin2}
\ea
where $\Theta_{\mu\nu}$ can be written explicitly for a suitable integral or series representation of the form factor \cite{Calcagni:2018lyd}.

Since the free equation of motion \Eq{F22fix} of the graviton is the same as the free massless version of the equation of motion for a scalar field with fractional d'Alembertian kinetic term, we expect the unitarity and renormalizability analysis carried out in \cite[section 4.1]{mf1} for the scalar field theory $T[\B^\g]$ to hold also for the gravitational theory defined by the action \Eq{act4fin2}. Concerning unitarity, $\g$ can take values only within the intervals
\be\label{rg1}
-2n<\g<1-2n\leq 1\,,\qquad n\in\mathbb{N}\,,
\ee
reduced to
\be\label{rg1min}
0<\g<1
\ee
if we insist on having a geometry with well-defined spectral dimension $\ds=D/\g$ \cite{mf1}. The spectral dimension is the dimensionality of spacetime felt by a diffusing probe particle, so that it is a meaningful geometric indicator only when it is positive semi-definite.

Regarding renormalizability, power-counting arguments have been shown not to hold \cite{mf1} and we have to check order by order in perturbation theory. At one loop, the fractional operator does not introduce divergences provided 
\be\label{rg3}
\g\neq \frac{D}{4}-\frac{n}{2}\,,\qquad n\in\mathbb{N}\,.
\ee
In four dimensions and in the above unitarity interval, it means that $\g\neq 1/2$.

Exponential or asymptotically polynomial form factors have been the main election in non-local quantum gravity, a unitary and super-renormalizable perturbative QFT of gravity \cite{Modesto:2011kw,BGKM,Modesto:2017sdr}. There, the action is \Eq{act4} and the form factor in \Eq{waveq} is $\cF_2\propto[\exp\H(\ell_*^2\B)-1]/\B$, where $\H(-z)$ is an entire function, so that the graviton equation of motion \Eq{waveq} is $\exp[-\H(\ell_*^2\B)]\B h_{\mu\nu}=0$ and no extra poles or branch cuts are introduced. Exponential form factors correspond to $\H(-z)=z^n$, $n=1,2,\dots$. Asymptotically polynomial form factors are more complicated but they all have the asymptotic limit $\exp\H(-z)\sim |z|^{n_{\rm deg}}$ for some integer power $n_{\rm deg}$. This limit is very similar to what we want to obtain in fractional gravity, the only difference being that instead of having $|z|^{n_{\rm deg}}$ asymptotically we have $z^{\g-1}$ from the start, where $\g$ is non-integer. There is also another similitude between the two theories. From the definition of the gamma function,
\ba
(m^2-\B)^\g&=&(m^2-\B)^{n}(m^2-\B)^{\g-n}\nonumber\\
&=&\frac{1}{\Gamma(n-\g)}\int_0^{+\infty}\rmd\tau\,\tau^{n-1-\g}\,(m^2-\B)^{n}\rme^{-\t(m^2-\B)}.\label{intpar}
\ea
In the sense of this integral representation, the fractional d'Alembertian is a sort of weighting of the exponential form factor with a fractional measure. Assuming $\g<1$,
\be\label{intpar20}
(-\B)^{\g-1}=\frac{1}{\Gamma(1-\g)}\int_0^{+\infty}\rmd\tau\,\tau^{-\g}\rme^{\tau\B}.
\ee
Although also our theory is non-local and quantum, we will keep a different naming to avoid confusion. Quantum gravity with fractional operators, or fractional gravity in short, goes one step further than non-local quantum gravity inasmuch as its non-local operators are non-analytic. In this sense, fractional gravity is more difficult and makes a more radical departure than non-local quantum gravity from standard QFT. Still, it is fascinating that we ended up with the same Lagrangian asymptotically in the UV, except for the value of the powers in the derivative operators (integer $n_{\rm deg}$ in asymptotically polynomial non-local quantum gravity, non-integer $\g$ in fractional gravity).

%%%%%%%%%%%%%%%%%%%%%%%%%%%%%%%%%%%%%%%%%%%%%%%%%%%%%%%%%%%%%%%%%%%%%%%%%%%%%

\subsection{Theory \texorpdfstring{$T[\B+\B^\g]$}{TBBg}}

The kinetic term of the graviton in this theory is a multi-fractional operator given by the composition of an ordinary and a fractional d'Alembertian (the relative sign between the two terms can also be negative but we do not discuss this point here):
\be\label{F21}
\cF_2(\B)=\ell_*^2(-\ell_*^2\B)^{\g-2}\qquad\Longrightarrow\qquad \left[\B+\ell_*^{-2}(-\ell_*^2\B)^\g\right] h_{\mu\nu}=0\,.
\ee
From the integral representation \Eq{intpar} with $\g<2$,
\be\label{intpar3}
\cF_2(\B)=\ell_*^{2(\g-1)}(-\B)^{\g-2}=\frac{\ell_*^{2(\g-1)}}{\Gamma(2-\g)}\int_0^{+\infty}\rmd\tau\,\tau^{1-\g}\rme^{\tau\B}.
\ee
Using Duhamel identity for the exponential of an operator $\cO$,
\be\label{paraH}
\de \rme^{\tau\cO} = \int_0^{\tau} \rmd q\, \rme^{q\cO}(\de \cO)\rme^{(\tau-q)\cO}\,,
\ee
for two generic symmetric rank-2 tensors $A_{\a\b}$ and $B^{\a\b}$ we have
\ba
\fl A_{\a\b}\de\cF_2B^{\a\b} &=& \frac{\ell_*^{2(\g-1)}}{\Gamma(2-\g)}\int_0^{+\infty}\rmd\tau\,\tau^{1-\g}A_{\a\b}\de\rme^{\tau\B}B^{\a\b}\nonumber\\
\fl &=& \frac{\ell_*^{2(\g-1)}}{\Gamma(2-\g)}\int_0^{+\infty}\rmd\tau\,\tau^{1-\g}\int_0^{\tau} \rmd q\, A_{\a\b}\rme^{q\B} (\de \B)\rme^{(\tau-q)\B}B^{\a\b}\nonumber\\
\fl &=& \frac{\ell_*^{2(\g-1)}}{\Gamma(2-\g)}\int_0^{+\infty}\rmd\tau\,\tau^{1-\g}\int_0^{\tau} \rmd q\, \rme^{q\B}A_{\a\b} (\de \B)\rme^{(\tau-q)\B}B^{\a\b}+O(\N)\,.\nonumber
\ea
Recalling that \cite{Calcagni:2018lyd}
%\bs
\ba
\fl A_{\a\b}(\de\B) B^{\a\b} &=& \de g^{\mu\nu}\bar\Theta_{\mu\nu}(A_{\a\b},B^{\a\b})+O(\N)\,,\label{dbmn}\\
\fl \bar\Theta_{\mu\nu}(A_{\a\b},B^{\a\b}) &:=& -\N_\mu A_{\a\b}\N_\nu B^{\a\b}+\frac14 g_{\mu\nu} \N_\rho(A_{\a\b}\N^\rho B^{\a\b}+B^{\a\b}\N^\rho A_{\a\b})\nonumber\\
\fl &&+\frac14 g_{\mu\nu} \N_\rho(A_{\a\b}\N^\rho B^{\a\b}-B^{\a\b}\N^\rho A_{\a\b})+\N_\a(A_{\mu\b}\N^\a B^\b_{\ \nu}-B^\b_{\ \nu}\N^\a A_{\mu\b})\nonumber\\
\fl && +\N_{\b}(B_{\mu\a}\N_\nu A^{\a\b}-A^{\a\b}\N_\nu B_{\mu\a})+\N_\a(A_{\mu\b}\N_\nu B^{\b\a}-B^{\b\a}\N_\nu A_{\mu\b})\,,\nonumber\\\label{barth}
\ea%\es
we get
\ba
\fl R_{\a\b}\de\cF_2G^{\a\b} &=& \frac{\ell_*^{2(\g-1)}}{\Gamma(2-\g)}\de g^{\mu\nu}\int_0^{+\infty}\rmd\tau\,\tau^{1-\g}\int_0^{\tau} \rmd q\,\bar\Theta_{\mu\nu}[\rme^{q\B}R_{\a\b},\rme^{(\tau-q)\B}G^{\a\b}]+O(\N)\nonumber\\
\fl &=& \de g^{\mu\nu}\Theta_{\mu\nu}(R_{\a\b},G^{\a\b})+O(\N)\,,
\ea
where
\be\label{thethefrac}
\fl \Theta_{\mu\nu}(R_{\a\b},G^{\a\b}) = \frac{\ell_*^{2(\g-1)}}{\Gamma(2-\g)}\int_0^{+\infty}\rmd\tau\,\tau^{1-\g}\int_0^{\tau} \rmd q\,\bar\Theta_{\mu\nu}[\rme^{q\B}R_{\a\b},\rme^{(\tau-q)\B}G^{\a\b}]
\ee
and $\bar\Theta_{\mu\nu}$ is second-order in derivatives. To summarize, for the theory with form factor \Eq{F21} the action \Eq{act4} simplifies to 
\be
\boxd{S=\frac{1}{2\k^2}\int\rmd^Dx\,\sqrt{|g|}\,\left[R-2\Lambda+\ell_*^2 G_{\mu\nu}(-\ell_*^2\B)^{\g-2}\,R^{\mu\nu}\right],\label{act4fin}}
\ee
while the equations of motion \Eq{EinEq1} read
\ba
\fl \k^2 T_{\mu\nu} &=& [1-(-\ell_*^2\B)^{\g-1}] G_{\mu\nu}+\Lambda g_{\mu\nu}\nonumber\\
\fl &&+\ell_*^2 g_{\mu\nu}\N^\s\N^\t(-\ell_*^2\B)^{\g-2} G_{\s\t} -2\ell_*^2\N^\s\N_{(\mu}(-\ell_*^2\B)^{\g-2} G_{\nu)\s}\nonumber\\
\fl &&-\frac{1}{2}\ell_*^2 g_{\mu\nu}G_{\s\t}(-\ell_*^2\B)^{\g-2} R^{\s\t}+2\ell_*^2G^\s_{\ (\mu} (-\ell_*^2\B)^{\g-2}  G_{\nu)\s}\nonumber\\
\fl &&+\frac12\ell_*^2[G_{\mu\nu}(-\ell_*^2\B)^{\g-2}  R+R(-\ell_*^2\B)^{\g-2} G_{\mu\nu}]+\Theta_{\mu\nu}(R_{\a\b},G^{\a\b})\,,\label{EinEq1fin}
\ea
with $\Theta_{\mu\nu}$ given by \Eq{thethefrac}.

At the quantum level, the theory should enjoy the same properties and problems of its scalar-field counterpart, due to the equivalence between the graviton linearized equation \Eq{F21} and the free massless version of the equation of motion for the scalar theory $T[\B+\B^\g]$ \cite[section 4.2]{mf1}. In this case, the results of \cite[section 4.1]{mf1} for the theory $T[\B^\g]$ do not apply to the UV of the theory $T[\B+\B^\g]$ because the fractional operator dominates at low energies/low curvature and at short scales the theory behaves as $T[\B]$, not $T[\B^\g]$. Therefore, the exclusion points \Eq{rg3} guarantee the absence of IR rather than UV divergences. Although a calculation with the full multi-scale kinetic term would say the final word, the theory is probably non-renormalizable just like Einstein gravity, since the $\B$ kinetic term dominates at short scales. Still, this theory can be used as an interesting generator of classical and quantum modifications of gravity at large scales, with cosmological applications such as in the problems of dark matter and dark energy (see section \ref{compa}).

%%%%%%%%%%%%%%%%%%%%%%%%%%%%%%%%%%%%%%%%%%%%%%%%%%%%%%%%%%%%%%%%%%%%%%%%%%%%%

\subsection{Theory \texorpdfstring{$T[\B^{\g(\ell)}]$}{TBgell}}

A graviton kinetic term given by a single fractional d'Alembertian with scale-dependent order is obtained from \Eq{F22fix} with $\g\to\g(\ell)$:
\be\label{F22}
\cF_2(\B)=\frac{(-\ell_*^2\B)^{\g(\ell)-1}-1}{\B}\,,\qquad (-\B)^{\g(\ell)}\, h_{\mu\nu}=0\,.
\ee
The action and equations of motion for the theory with such form factor can be made explicit in a similar way and are 
\be
\fl \boxd{S=\frac{1}{2\k^2\ell_*}\int_0^{+\infty}\rmd\ell\,\tau(\ell)\int\rmd^Dx\,\sqrt{|g|}\,\left[R-2\Lambda+ G_{\mu\nu}\frac{(-\ell_*^2\B)^{\g(\ell)-1}-1}{\B}\,R^{\mu\nu}\right],\label{act4fin23}}
\ee
and \Eq{EinEq1fin2} with $\g\to\g(\ell)$, where we chose the weight $\tau$ such that $\int_0^{+\infty}\rmd\ell\,\tau(\ell)=\ell_*$. 
%\ba
%\k^2 T_{\mu\nu} &=& (-\ell_*^2\B)^{\g(\ell)-1} G_{\mu\nu}+\Lambda g_{\mu\nu}\nonumber\\
%&&+g_{\mu\nu}\N^\s\N^\t\frac{(-\ell_*^2\B)^{\g(\ell)-1}-1}{\B} G_{\s\t} -2\N^\s\N_{(\mu}\frac{(-\ell_*^2\B)^{\g(\ell)-1}-1}{\B} G_{\nu)\s}\nonumber\\
%&&-\frac{1}{2} g_{\mu\nu}G_{\s\t}\frac{(-\ell_*^2\B)^{\g(\ell)-1}-1}{\B} R^{\s\t}+2G^\s_{\ (\mu} \frac{(-\ell_*^2\B)^{\g(\ell)-1}-1}{\B} G_{\nu)\s}\nonumber\\
%&&+\frac12\left[G_{\mu\nu}\frac{(-\ell_*^2\B)^{\g(\ell)-1}-1}{\B}  R+R\frac{(-\ell_*^2\B)^{\g(\ell)-1}-1}{\B} G_{\mu\nu}\right]+\Theta_{\mu\nu}(R_{\a\b},G^{\a\b})\,,\nonumber\\
%\label{EinEq1fin2}
%\ea
 As we discussed in section \ref{tigall}, the equations of motion are valid at any given $\ell$.

The case with scale-dependent $\g(\ell)$ represented by the form factor and graviton equation \Eq{F22} may be more promising as a fundamental theory than the theories $T[\B^\g]$ (unitary and one-loop finite but with no multi-scaling) and $T[\B+\B^\g]$ (never unitary and renormalizable at the same time). The unitarity and one-loop finiteness set $\g(\ell)\in (0,1/2)\cup(1/2,1)$ corresponds to a class of theories well behaved in the UV, as originally desired, while maintaining the possibility to find a non-trivial cosmological imprint.

%%%%%%%%%%%%%%%%%%%%%%%%%%%%%%%%%%%%%%%%%%%%%%%%%%%%%%%%%%%%%%%%%%%%%%%%%%%%%
%%%%%%%%%%%%%%%%%%%%%%%%%%%%%%%%%%%%%%%%%%%%%%%%%%%%%%%%%%%%%%%%%%%%%%%%%%%%%

\section{Comparison with the literature}\label{compa}

As an early mathematical utilization of fractional calculus to gravity, applying fractional integrals on the ordinary Poisson equation one can relate the pointwise and semi-infinite linear mass distributions by continuously deforming one into the other \cite{RMRW}. However, this is not a model of fractional gravity where fractional integrals or derivatives are part of the gravity-matter interweave, which is what we will review in this section. Also, a source of confusion in the literature may be that models called `fractional' indiscriminately refer to modifications of the integral and/or the differential structure of gravitational dynamics. Here we strictly refer to models with fractional derivatives, not with fractional integration measure, which are reviewed in \cite[section 1.2]{frc11}.

\subsection{Limited portions of fractional gravity}

Several models studied two limited aspects of gravity with fractional derivatives without embedding them in a fundamental theory: Newtonian gravity and cosmology.
\begin{itemize}
\item The linearized equations of motion and some general properties of ghost-free large-distance modifications of gravity were studied in \cite{Dva06}. Here the graviton kinetic term is augmented by a fractional d'Alembertian, so that, up to other trace $h_\mu^\mu$ and second-order derivative terms, $[\B+ r_{\rm c}^{2(\g-1)}\B^\g] h_{\mu\nu}\sim \k^2 T_{\mu\nu}$, where $r_{\rm c}$ is a cosmological scale and $\g<1$. Using the spectral representation reviewed and employed in \cite{mf1}, the unitarity constraint $\g\geq 0$ was evinced. The resulting range $0<\g<1$ corresponds to the unitarity constraint \Eq{rg1min}. No non-linear gravitational action was proposed.
\item Newton's potential was derived from an \emph{ad hoc} fractional Poisson equation \cite{MBRa,Varieschi:2017xjc,Giusti:2020rul,Giusti:2020kcv}. As an application, under the hypothesis that the matter distribution of galaxies behaves as a fractal medium with non-integer dimension, solving a Poisson equation with fractional Laplacian $(-\Delta)^\g$ one can describe the properties of such matter distribution with a fractional version of Newtonian gravity and account for the observed galaxy rotation curves without invoking dark matter, if the fractional exponent $\g$ is close to $3/2$ \cite{Giusti:2020rul,Giusti:2020kcv}.\footnote{A similar result can be obtained from a Poisson equation originated from a Gauss theorem with fractional measure \cite{Varieschi:2020ioh,Varieschi:2020dnd,Varieschi:2020hvp}. The ensuing Laplacian operator is made of ordinary derivatives but has anomalous dimension and is the one appearing in the diffusion equations proposed for fractal media \cite{OSP,GiR,MGN}. This model falls into the category of scenarios mentioned at the beginning of section \ref{compa} that we do not consider here but, nevertheless, we notice that a theory that could easily accommodate Varieschi's model and extend it to a matter distribution with scale-dependent dimension could be any of the multi-fractional theories $T_1$, $T_v$ or $T_q$ with an ordinary differential structure \cite{revmu,Calcagni:2021ipd,fra1,fra2,fra3}.}
\item The ordinary time derivatives of the Friedmann equations of homogeneous and isotropic cosmology were replaced by fractional derivatives in \cite{ElN05,Roberts:2009ix,Shchigolev:2010vh,Barrientos:2020kfp}. In particular, in \cite{Barrientos:2020kfp} it was shown that supernov\ae\ data on the late-time acceleration of the universe can be explained by fractional Friedmann equations with derivative order close to $3/2$. These equations were not obtained from the symmetry reduction of any background-independent non-linear gravitational action.
\item Another attempt to explain dark energy was made using thermodynamical arguments based on a Schr\"odinger equation with fractional Laplacian with $\g=-3/2$ \cite{Landim:2021www}.
\end{itemize}
It would be important to check whether these phenomenological results, interesting \emph{per se} but lacking a robust theoretical motivation, can be obtained in the gravitational theories proposed here, without violating experimental constraints on gravity at sub-galactic and cosmological scales.

\subsection{Full non-linear theories of fractional gravity}

Apart from the class of multi-fractional theories $T_\g$, there are not many other proposals for a non-linear action of gravity with fractional operators.
\begin{itemize} 
\item A non-minimal scalar-tensor theory with fractional Laplacians $\Delta^{1/2}$ and $\Delta^{3/2}$ was formulated in order to preserve detailed balance in the matter sector in the first formulation of Ho\v{r}ava--Lifshitz gravity \cite{Calcagni:2009qw}. Detailed balance is a condition imposed on the action that suppresses the proliferation of operators at the quantum level. The ensuing action is complicated and is made of operators of order 6 in spatial derivatives. It contains covariant derivatives of the spatial Ricci tensor $R_{ij}$ and of a scalar $\phi$, as well as the fractional terms $\Delta^{1/2}\phi$, $\Delta^{3/2}\phi$ and $\Delta^{1/2} R_{ij}$. The theory is unitary and is argued to be renormalizable but it has other problems because Lorentz invariance is not recovered in the IR \cite{Calcagni:2009qw}.
\item The generalization of the Ricci tensor and the Einstein equations to fractional derivatives was written down in \cite{Munkhammar:2010gq}, where the fractional Poisson equation was also derived, thus justifying the starting point of \cite{MBRa,Varieschi:2017xjc} one step further. However, the modified Einstein equations \cite{Munkhammar:2010gq}
\be\label{eomuk}
R_{\mu\nu}^{(\g)}-\frac12 g_{\mu\nu}R^{(\g)}=\k^2T_{\mu\nu}
\ee
were not derived from the variation of an action. In \cite{Munkhammar:2010gq}, the Ricci tensor is \Eq{ricg} with fractional Levi-Civita connection \Eq{leci}, where the mixed fractional derivative $\cD_\mu^\g$ is replaced everywhere by the Riemann--Liouville fractional derivative ${}_\textsc{rl}\p_\mu^\g$ \cite{SKM,KST}. Although the Ricci tensor looks similar, the dynamics is not because the derivative operator is different and, in particular, the Riemann--Liouville derivative of a constant is not zero. This can create serious problems when defining local inertial frames, the role of Minkowski metric, the covariant conservation of the metric, and so on \cite{frc1}. Also, the expressions of the Levi-Civita connection and Ricci tensor were used in \cite{Munkhammar:2010gq} to define the equations of motion \Eq{eomuk}, while in the case of the theory $T[\p^\g]$ we have built the Levi-Civita connection and Ricci tensor according to the paradigm of multi-scale spacetimes, then we defined the action \Eq{actg0}, then we derived the equations of motion \Eq{eomukfix}, which include the contribution of the operator $\cO_{\mu\nu}$ missing in \Eq{eomuk}. Also, in our case we regard $T[\p^\g]$ as a sort of toy model for the multi-fractional versions of the theory, unless $\g\approx 1$.
\item The same Ricci tensor and Einstein equations of the previous bullet were constructed by Vacaru in \cite{Vacaru:2010wn} using instead Caputo fractional derivative $\p_\mu^\g$. This proposal is much more rigorous than the previous one since it relies on the formalism of Finsler geometry \cite{Vacaru:2018yha} and, in particular, a structure of fractional manifold built with fractional differentials similar to those later employed in \cite{frc1}. Once again, the Einstein equations were not derived from an action, although this should be possible using the fractional Euler--Lagrange equations found in \cite{Baleanu:2010gn,Baleanu:2010wc}. A preliminary attempt to quantization was sketched in \cite{Baleanu:2010gq} but at a very formal level. Overall, this theory is more a subject of mathematical physics than of observation-oriented QFT or quantum gravity.
\end{itemize}

\subsection{Other theories of quantum or non-local gravity}\label{azas}

Finally, we compare the theories with multi-fractional operators in the class $T_\g$ with other quantum gravities with dimensional flow, i.e., where the dimension of spacetime changes with the probed scale. The choice of the ordinary spacetime measure $\rmd^Dx$ draws the theory closer to those quantum gravities, most of them summarized in \cite{Calcagni:2019ngc}, where the spectral dimension $\ds$ of spacetime (governed by the type of kinetic term) varies with the probed scale while the Hausdorff dimension $\dh$ (governed by the spacetime measure) does not \cite{mf1}. String field theory, asymptotic safety, causal dynamical triangulations (CDT), non-local quantum gravity, most non-commutative spacetimes, Ho\v{r}ava--Lifshitz gravity and a few others fall into this category ($\dh=D$); notable exceptions, where also $\dh$ varies, are cyclic-invariant theories on $\k$-Minkowski spacetime and the set of mutually related discretized theories made of group field theory (GFT), spin foams and loop quantum gravity (LQG).

In \cite{mf1}, it was found that the spectral dimension of spacetime in the theories with multi-fractional operators is asymptotically given by the ratio between the topological dimension $D$ ($=4$ in physical scenarios) and the power of the d'Alembertian, either 1 or the fractional order $\g$, depending on which operator between $\B$ and $\B^\g$ dominates at the scale at which one is calculating $\ds$. For the theories $T[\p+\p^\g]$ and $T[\B+\B^\g]$, the operator of order $\g$ dominates at short scales if $\g>1$ and at large scales if $\g<1$, while in the theories $T[\p^{\g(\ell)}]$ and $T[\B^{\g(\ell)}]$ the dimensional flow of $\ds$ is more free because the profile $\g(\ell)$ can be constructed \emph{ad hoc}. In this case, one can always consider geometries where $\ds=D/\g$ at small scales where $\g\neq 1$ is the asymptotic value of $\g(\ell)$. We consider this more general possibility here.

When $\g>1$, the spectral dimension decreases from $\ds\simeq D$ at large scales to 
\be\label{dsas}
\ds\simeq \frac{D}{\g}
\ee
at short scales, as in Stelle gravity \cite{CMNa}, string field theory \cite{ACEMN,CaMo1}, asymptotic safety \cite{LaR5}, CDT \cite{CoJu}, non-local quantum gravity \cite{Modesto:2011kw}, Ho\v{r}ava--Lifshitz gravity \cite{Hor3} and GFT/spin foams/LQG \cite{ACAP,COT3,MiTr}. Therefore, combining constancy of $\dh$ and the decrease of $\ds$ towards the UV, the theories with most similar dimensional flow with respect to ours are string field theory, asymptotic safety, CDT, non-local quantum gravity and Ho\v{r}ava--Lifshitz gravity, among others. A special case is the limit $\g\to+\infty$ for all directions $\mu$. The spectral dimension vanishes in the UV, as in non-local quantum gravity  and in the discreteness-effects scale range of GFT/spin foams/LQG.

When $\g<1$, the spectral dimension increases to \Eq{dsas} at small scales as in $\k$-Minkowski spacetime with bicross-product Laplacian \cite{ArTr,Eckstein:2020gjd}, for which $\ds^\textsc{uv}=6$. To have \Eq{dsas} reproduce this value in $D=4$ dimensions, it should be $\g=2/3$. 

In the limit $\g\to 0^+$, one would approximate the dimensional flow of $\k$-Minkowski spacetime with relative-locality Laplacian \cite{ArTr,Eckstein:2020gjd} as well as Padmanabhan's model of non-local black holes \cite{ArCa1}, where $\ds$ diverges in the UV. 

In the same limit $\g\to 0^+$, the action \Eq{act4fin} of the theory $T[\B+\B^\g]$ reproduces, up to a Ricci tensor-Ricci tensor term, the action of a phenomenological non-local model of IR modifications of gravity, dubbed RR model \cite{Maggiore:2014sia}:
\be\label{Calctolag2}
\cL=R-\frac{m^2}{6}R\frac{1}{\B^2}\,R\,.
\ee
This model, which is not of quantum gravity, was proposed to explain dark energy and to leave a possibly strong cosmological imprint in the propagation of gravitational waves, but it is ruled out because it violates the bounds on the time variation of the effective Newton's constant \cite{Belgacem:2020pdz}.

It is noteworthy that the theory \Eq{act4fin} also recovers the models of IR non-local gravity with the Ricci tensor: in the limit $\g\to 0^+$, the $R_{\mu\nu}\B^{-2}R^{\mu\nu}$ model \cite{Cusin:2015rex,Zhang:2016ykx} augmented with the above $RR$ term; in the limit $\g\to 1^-$, 
the $R_{\mu\nu}\B^{-1}R^{\mu\nu}$ model \cite{Ferreira:2013tqn,Nersisyan:2016jta} augmented with an $RR$ term or, exactly, the model \cite{Barvinsky:2003kg,Barvinsky:2005db,Barvinsky:2011hd,Barvinsky:2011rk} 
\be\label{Calctolag3}
\cL=R- R_{\mu\nu}\frac{1}{\B}\,G^{\mu\nu}\,.
\ee
Choosing $\g$ close but not equal to 0 or 1 within the range allowed by unitarity, it might be possible to obtain a phenomenology similar to these models without the instability problems affecting some of them \cite{Nersisyan:2016jta}.

%%%%%%%%%%%%%%%%%%%%%%%%%%%%%%%%%%%%%%%%%%%%%%%%%%%%%%%%%%%%%%%%%%%%%%%%%%%%%
%%%%%%%%%%%%%%%%%%%%%%%%%%%%%%%%%%%%%%%%%%%%%%%%%%%%%%%%%%%%%%%%%%%%%%%%%%%%%

\section{Conclusions and perspective}\label{conc}

In this paper, we studied the classical and quantum properties of gravitational theories with fractional kinetic terms, respecting or violating Lorentz invariance. The equations of motion are non-local and include extra terms with respect to the ordinary Einstein equations. As in the scalar-field case \cite{mf1}, the theories $T[\p^\g]$, $T[\p+\p^\g]$ and $T[\p^{\g(\ell)}]$ with fractional derivatives are more difficult and less manageable than their covariant counterparts $T[\B^\g]$, $T[\B+\B^\g]$ and $T[\B^{\g(\ell)}]$ with fractional d'Alembertian operators. The latter, however, cannot be at the same time unitary and power-counting renormalizable but there are cases where unitarity and one-loop finiteness can be realized simultaneously. Although this is actually impossible when the kinetic operator is the sum of integer and fractional d'Alembertians as in $T[\B+\B^\g]$, the quantum theory displays an interesting IR finiteness and large-scales modifications with, hopefully, cosmological applications.

The agenda for the future is filled with many items. On the theoretical side, fractional gravitational theories need a full exploration ranging from classical (e.g., cosmological) solutions to perturbative renormalizability, in the cases where this is possible. Early- and late-time cosmology would deserve to be studied, with emphasis on dark energy. We have seen in \Eq{Calctolag2} and \Eq{Calctolag3} that, in the limits $\g\to 0^+$ and $\g\to 1^-$, the theory $T[\B+\B^\g]$ resembles some realizations of IR non-local gravity that can generate late-time acceleration \cite{Belgacem:2020pdz}. We do not know whether the fact that $\g$ is not exactly 0 or 1 can evade the Lunar Laser Ranging bound violated by the RR model or the instabilities displayed in other models of non-local IR gravity. Nevertheless, multi-scale dynamics with dimensional flow is known to sustain cosmological acceleration in all the other multi-fractional theories, either at early or at late times, or both \cite{revmu,Calcagni:2020ads}. Therefore, we expect the theories with fractional operators to have similar phenomenology.

There is still much work to do to understand the quantum properties of the class of theories $T_\g$. The failure of power-counting renormalizability as a tool to estimate divergences and the question about perturbative renormalizability are non-trivial and expose the added difficulties of fractional operators with respect to analytic non-local quantum gravity \cite{MoRa1,MoRa2}. In higher-order derivative theories (integer $\g$), we know that divergences only scale as powers of the cut-off energy $\Lambda_{\rm UV}$ and divergence counting can be done analytically. If $\g$ is rational, divergences become rational powers of $\Lambda_{\rm UV}$ and one deals with Puiseux series (series of powers with negative and fractional exponents), in which case one could still show convergence analytically. In fact, sub-leading divergences can be expressed by a finite multiplicity of $\Lambda_{\rm UV}$ powers. We already calculated one-loop diagrams for the scalar QFT $T[\B^\g]$ with analytic methods, even for irrational $\g$ \cite{mf1}. From two loops on, one might have to give up analytic methods and recur to numerical methods.

%When $\g$ is irrational, we expect a whole continuum of sub-leading divergences in the cut-off down to the logarithmic one. This may be problematic for the computation of higher-order loop integrals because, in order to regularize them, one should subtract all (also sub-leading) divergences. With a continuous spectrum, we cannot use standard subtraction schemes and we might have to resort to techniques beyond conventional renormalization in quantum field theory. This does not mean that perturbation theory is ill defined, but that it may not be so straightforward.

It would be interesting to explore more in detail the existence of a duality between the fractional theory $T[\B^\g]$ (single fractional d'Alembertian) and the multi-fractional theory $T_1$ (normal derivatives but multi-scale measure), as mentioned in section \ref{compa}. Speaking about these two scenarios, they have the potential of offering a robust embedding for the promising galaxy-rotation-curve models, alternative to dark matter, proposed by Giusti \cite{Giusti:2020rul,Giusti:2020kcv} (anomalous Poisson equation with fractional Laplacian, possibly related to the theory $T[\B^\g]$ or, more realistically, $T[\B+\B^\g]$) and Varieschi \cite{Varieschi:2020ioh,Varieschi:2020dnd,Varieschi:2020hvp} (anomalous Poisson equation with ordinary derivatives, possibly related to the theory $T_1$). If a duality between these theories actually existed, it could also explain why the modified Newtonian potential in the models by Giusti and Varieschi is so similar in some range in the effective dimension of matter distribution \cite{Varieschi:2020ioh}.

%%%%%%%%%%%%%%%%%%%%%%%%%%%%%%%%%%%%%%%%%%%%%%%%%%%%%%%%%%%%%%%%%%%%%%%%%%%%%
%%%%%%%%%%%%%%%%%%%%%%%%%%%%%%%%%%%%%%%%%%%%%%%%%%%%%%%%%%%%%%%%%%%%%%%%%%%%%

\section*{Acknowledgments}

The author is supported by the I+D grants FIS2017-86497-C2-2-P and PID2020-118159GB-C41 of the Spanish Ministry of Science and Innovation. He thanks F Briscese, L Modesto, L Rachwa\l\ and especially G Nardelli for useful comments.

%%%%%%%%%%%%%%%%%%%%%%%%%%%%%%%%%%%%%%%%%%%%%%%%%%%%%%%%%%%%%%%%%%%%%%%%%%%%%
%%%%%%%%%%%%%%%%%%%%%%%%%%%%%%%%%%%%%%%%%%%%%%%%%%%%%%%%%%%%%%%%%%%%%%%%%%%%%

\end{document}